\documentclass{article}

\usepackage{geometry}
\usepackage{latexsym, amsmath, amsfonts, amssymb, dsfont, subfigure, graphicx, natbib, url, booktabs, multirow}

\newcommand{\Lc}{\ensuremath{\mathcal{L}}}

\newcommand{\Ed}{\ensuremath{\mathds{E}}}
\newcommand{\Nd}{\ensuremath{\mathds{N}}}
\newcommand{\Pd}{\ensuremath{\mathds{P}}}
\newcommand{\Rd}{\ensuremath{\mathds{R}}}

\newcommand{\set}[1]{ \ensuremath{\left\{ #1 \right\}} }

\begin{document}
%\begin{frontmatter}

\title{A Coupled Markov Chain Approach to Credit Risk Modeling\footnote{Research partly supported by the Austrian National Bank Jubil{\"a}umsfond Research Grant 12306.}}

\author{David Wozabal\footnote{Department of Business Administration, University of Vienna, A-1210 Vienna, Br\"unner Stra\ss e 72} \and Ronald Hochreiter\footnote{Department of Finance, Accounting and Statistics. WU Vienna University of Economics and Business, A-1090 Vienna, Augasse 2-6}}
\date{}

\maketitle

\begin{abstract}
We propose a Markov chain model for credit rating changes. We do not use any distributional assumptions on the asset values of the rated companies but directly model the rating transitions process. The parameters of the model are estimated by a maximum likelihood approach using historical rating transitions and heuristic global optimization techniques.

We benchmark the model against a GLMM model in the context of bond portfolio risk management. The proposed model yields stronger dependencies and higher risks than the GLMM model. As a result, the risk optimal portfolios are more conservative than the decisions resulting from the benchmark model.
\end{abstract}
%\begin{keyword}
%Credit Risk \sep Markov Models \sep Ratings \sep Conditional Value-at-Risk \sep Bond portfolios

%\emph{JEL:} G11 \sep G01 \sep C44 \sep C14 \sep D81
%\end{keyword}

%\end{frontmatter}
%\newpage

%%%%%%%%%%%%%%%%%%%%%%%%%%%
% SECTION 1: Introduction %
%%%%%%%%%%%%%%%%%%%%%%%%%%%
\section{Introduction}
In this paper, we present a coupled Markov chain (CMC) model which builds on the approach in \citet{KP07}. The aim of the model is to come up with a statistical description of the joint probabilities of credit rating changes of companies, which does not depend on distributional assumptions of the joint distribution of the asset values of the companies. We assume the that the individual rating transitions follow Markov processes and model the dependency between rating migrations of different companies by coupling the corresponding Markov chains. The advantage of being able to describe the dependencies of the credit quality of multiple debtors is that risk management on a portfolio level can be based on such a model. Therefore, we assess the quality of the model in the context of a stylized bond portfolio optimization problem and compare the portfolio decisions based on the proposed coupled Markov model with the decisions based on a GLMM model from the literature. The results show that the proposed model yields more conservative decisions then the GLMM model.

A major advantage of the proposed model over \cite{KP07} is that it lends itself to statistical estimation of the parameters. More specifically, we derive the likelihood function of the model and develop methods for finding solutions to the maximum likelihood problem -- a task which is complicated by the fact that the likelihood function is non-convex and computationally expensive to evaluate.

Although agency ratings have been criticized for their sluggish response to fast evolving events (see \citet{Altm98, CrBo02, LaSk02, NiPe00}), many models use credit ratings as a basis for assessing credit risk. The credit rating of a company condenses a range of qualitative and quantitative assessments of the credit worthiness of a company and therefore is a signal for the credit quality of the debtor, which is consistent over time as well as among different debtors. Furthermore, rating based valuations are of increasing importance since pending new banking regulations use ratings as an important input for calculating capital requirements for banks (see \citet{BIS05}).

The most commonly used rating based method for modeling credit risk is the \emph{CreditMetrics} approach. The main idea behind \emph{CreditMetrics} is similar to the one in this paper: the current rating of a company influences the default probability in the next period. The difference, however, to the proposed approach is the copula used to specify the joint behavior of the rating processes for different companies. In the \emph{CreditMetrics} approach, a Gaussian copula is used for this purpose. There exist a range of models describing the joint default behavior of companies in the literature: for excellent surveys and model classifications see for example \citet{CrGaMa00, DuffieSingleton2003, FrMc2003a, FrSc08, Gord00} or \citet{QRM}.

In \cite{McWe06, McWe07}, the authors propose a generalized linear mixed model (GLMM) for rating transitions which is estimated using Bayesian techniques. The model describes systematic risk factors as a combination of fixed and random effects and also allows for serial correlations in the unobserved risk factors and hence for so called \emph{rating momentum} on a macroeconomic scale.

\cite{StTuTu09} propose a model for continuous \emph{credit worthiness} variables, which are translated to discrete ratings by identifying a rating class with an interval of the credit worthiness score. The continuous credit worthiness variables are allowed to depend on obligor specific as well as macroeconomic factors, whereby the latter are used for modeling dependencies in rating transitions of different obligors.

In \cite{KoEl08} credit quality is modeled by a hidden Markov model. The published credit ratings are considered to be noisy signals that give an indication of the true credit worthiness.

Models for credit quality based on ratings are also frequently used in the pricing and risk management literature, see for example \cite{JaLaSt97, Kiji98a, Kiji98b}.

Note that there is some empirical evidence hinting to the fact that the Markov assumption of credit ratings does not always hold (see \citet{Altm98, LaSk02, NiPe00}). The reasons for this might be contagion effects (cf. \cite{GiWe06}) or long range dependencies in macroeconomic variables. Nevertheless, we do not consider more complicated models as the Markov assumption does not seem to be \emph{too wrong} as shown in \citet{KiLa06} and is implicit in most credit risk models.

This paper is structured as follows: Section \ref{sec:model} is devoted to a discussion of the coupled Markov chain model. In Section \ref{sec:ml}, we discuss a maximum likelihood approach which is subsequently used to estimate the parameters of the model from empirical data in Section \ref{sec:estimation}. In Section \ref{sec:comparison}, we compare the proposed model to a model by \cite{McWe06} and discuss the differences of two models in a risk management context. Section \ref{sec:conclusion} concludes the paper.

%%%%%%%%%%%%%%%%%%%%%%%%
% SECTION 2: The Model %
%%%%%%%%%%%%%%%%%%%%%%%%
\section{The Model}
\label{sec:model}
The model is based on the ideas presented in \citet{KP07}. For sake of clarity, we postpone the discussion of the differences to the aforementioned paper to the end of this section.

We model joint rating transitions of companies in different rating classes belonging to different industry sectors, such that
\begin{enumerate}
	\item migrations of companies having the same credit rating are dependent;
	\item evolution of companies through credit ratings are dependent;
	\item every individual migration is governed by a Markovian matrix, which is the same for all the companies.
\end{enumerate}
In line with \cite{KP07}, we assume that the rating migration process of each company is Markov with identical rating transition matrix and that these processes are coupled in such a way that they are statistically dependent because of their dependence on common systematic factors.

We start by considering a diversified portfolio consisting of debt obligations of different firms $n \in \set{1, \ldots, N}$. The debtors are non-homogeneous in their credit ratings and belong to different industry sectors. Assume that there are $M$ non-default rating classes. The ratings are numbered in a descending order so that $1$ corresponds to the highest credit quality, while $M$ is next to the default class. For example, in terms of the rating scheme of Standard and Poor's (S\&P) we have, $1\leftrightarrow AAA$, $2\leftrightarrow AA$, $3\leftrightarrow A$, $4\leftrightarrow BBB$, $5\leftrightarrow BB$, $6\leftrightarrow B$, $7\leftrightarrow CCC$, $8\leftrightarrow CC$, $9\leftrightarrow C$ and $10\leftrightarrow D$ with $M=9$.

A company $n$ is fully characterized by its rating $x_n^t$ at time $t$ as well as its industry sector $s(n)$. Denote by $S \in \Nd$ the number of different industry sectors in the model, i.e. $s(n) \in \set{1,\ldots, S}$ for all $n \in \set{1, \ldots, N}$. Note that the sectoral classification can be replaced by any arbitrary discrete classification scheme without structurally changing the model. Possible alternative classifications could, for example, include size or geographic origin of the company.

Let $p_{i,j}$ be the probability that a company in rating class $i$ at the beginning of a time period ends up in rating class $j$ at the end of the period. In particular, $p_{i,M+1}$ is the probability that a debtor who is currently rated with $i$-th credit rating defaults within the next period. The $M\times (M+1)$ transition matrix $P=(p_{i,j})$ can be estimated using one of the various techniques proposed in literature and is also reported by the rating agencies themselves. Since we are mainly concerned with the coupling of the rating processes, we assume the matrix $P$ to be known.

We suppose the evolution of the portfolio is modeled by a multi-dimensional random process $X^t=(X_1^t, \dots, X_N^t)$. $X_n^t$ indicates the rating of company $n$ at the end of period $t$, whereby period $t$ is the timespan between time points $t-1$ and $t$. The marginals $X_n^t$ are modeled as dependent discrete-time Markov chains with state space $\{ 1, 2, \dots, M+1 \} $, transition probability matrix $P$, and absorbing state $M+1$.

Similar to the classical one factor Gaussian model, the starting point of the specification of the dependencies between the rating processes of individual companies is the decomposition of risk factors in an idiosyncratic and a systematic factor. We split the rating movement of company $n$ at time $t$ in two components: a systematic component $\eta_n^t$ and an idiosyncratic component $\xi_n^t$. Both components take values in \set{1, \ldots, M+1} with transition probabilities depending on $X_n^{t-1}$. The non-trivial joint distribution of the vector $(\eta_1^t, \ldots, \eta_N^t)$ is used to model dependency between the rating migration processes.

The rating of company $n$ at time $t$ is given by
\begin{equation} \label{eq:model}
X_n^t = \delta_n^t \xi_n^t + (1-\delta_n^t) \eta_n^t.
\end{equation}
The mixing between the idiosyncratic and the systematic component is achieved via a Bernoulli variable $\delta_n^t$ depending on the rating class of company $n$ at time $t-1$, $X_n^{t-1}$, as well as the industry sector $s(n)$ of company $n$. We define
\begin{equation}
q_{m,s}:=\Pd(\delta_n^t = 1 \mid X_n^{t-1} = m, s(n) = s), \quad \forall m:1\leq m \leq M, \; \forall s:1\leq s \leq S
\end{equation}
and $Q=(q_{m,s})_{1\leq m \leq M, 1\leq s \leq S}$.

While the variables $\delta_n^t$ and $\xi_n^t$ are independent of all the other variables, the variables $\eta_n^t$ are independent of $\delta_n^t$ and $\xi_n^t$, but have a non-trivial dependency structure within a given time period. The marginal distribution of $\xi_n^t$ and $\eta_n^t$ is dependent on the current rating class of the company $n$ and the transition matrix $P$. In particular, we define
\begin{equation}\label{eq:etaProb}
\Pd( \xi_n^t = i) = p_{X_n^{t-1}, i}, \quad \forall i \in \set{1, \ldots, M+1}, \; \forall n \in \set{1, \ldots, N}.
\end{equation}

To model the dependencies between the components of the random vector $\eta^t = (\eta_1^t, \ldots, \eta_N^t)$, we divide the transition of $\eta_n^t$ in two parts: its \textit{tendency} (i.e. up or down) and the \emph{magnitude} of the change. To be more specific, let $\chi^t=(\chi_1^t, \ldots, \chi_M^t)$ be a vector of Bernoulli variables, where $\chi_m^t$ determines whether a non-deteriorating move takes place for all companies $n$ with $X_n^{t-1} = m$ at time $t$, i.e.
\begin{equation}
\chi^t_m=\begin{cases} 1, & \eta^t_n\leq X_n^{t-1}, \; \forall n: X_n^{t-1} = m\\
0, & \eta^t_n > X_n^{t-1}, \; \forall n: X_n^{t-1} = m.
\end{cases}
\end{equation}
The probability of success of $\chi_m^t$ is given by $\Pd (\chi_m^t=1)=p_m^{+}:= \sum_{j=1}^{m}p_{m,j}.$
%\begin{equation}
%\Pd (\chi_m^t=1)=p_m^{+}:= \sum_{j=1}^{m}p_{m,j}.
%\end{equation}

Observe that there are not $N$ but only $M$ tendency variables $\chi_m^t$ per time period $t$. This renders the variables $\eta_n^t$ dependent via their dependance on $\chi^t$ and the non-trivial joint distribution of the same. Conditioned on the tendency, the magnitude of the change for a company $n$ with $X_n^{t-1} = m$ follows the distribution
\begin{equation}
\Pd (\eta_n^t=j \mid \chi_m^t=1)=\left\{
\begin{array}{ll}
p_{m,j}/p_m^{+}, & j\leq m, \\
0, & j>m
\end{array}
\right.
\end{equation}
and
\begin{equation}
\Pd (\eta_n^t=j \mid\chi_m^t=0 )=\left\{
\begin{array}{ll}
p_{m,j}/(1-p_m^{+}), & j>m \\
0, & j\leq m.
\end{array}
\right.
\end{equation}
Given the tendencies, the distribution of the magnitude is completely determined by the marginal distribution of the rating transitions, i.e. the matrix $P$. Note that unconditional distribution of $\eta_n^t$ with $X_n^{t-1} = m$ is
\begin{equation}
\Pd(\eta_n^t = j) = p_{m,j}, \quad \forall j: 1\leq j \leq M+1,
\end{equation}
i.e. identical to the distribution of $\xi_n^t$.

The decomposition of a rating move in a \emph{tendency} and a \emph{magnitude} is useful since the tendency part can be used to model the dependencies between different debtors via the joint distribution of $\chi^t = (\chi_1^t, \ldots, \chi_m^t)$, which we do not restrict in our model. Finally, variables $\eta_n^t \mid \chi_{X_n^{t-1}}^t$ are modeled independent of one another.

Since $X^t_n$ depends only on the variables $\delta_n^t$, $\xi_n^t$ and $\eta_n^t$ and these variables in turn only depend on $s(n)$ and $X_n^{t-1}$, $(X^t)_{t\geq 0}$ is a Markov chain, i.e.
\begin{eqnarray}
\Pd(X^t = x^t| X^{t-1}, \ldots, X^1) = \Pd(X^t = x^t| X^{t-1})
\end{eqnarray}
where $X^t = (X_1^t, \ldots, X_N^t)$ and $x^t \in \set{1, \ldots, M+1}^N$.

The model in \eqref{eq:model} differs from the model presented in \cite{KP07} (the KP model) in two aspects: in the specification of the dependencies and in the way parameter estimates are obtained. We start with discussing the former: the KP model reads
\begin{equation} \label{eq:KPModel}
X_n^t = \delta_n^t \xi_n^t + (1-\delta_n^t)\eta^t_{X_n^{t-1}}.
\end{equation}
Note that at every point in time $t$, there are only $(M+1)$ different variables $\eta^t$ and not $N$ of them like in the formulation proposed in \eqref{eq:model}. This implies that all companies $n$ with rating class $i$, whose rating move is determined by the systematic risk factors (i.e. $\delta_n^t = 0$) move exactly to the same rating class in period $t$. Since this seems to be a strong assumption, we allow that companies are affected by the common risk factors to various degrees: while the direction remains same for all the companies in a specific rating class, the magnitude of the effect of the common risk factors is allowed to vary. This additional degree of freedom makes the filtering problem posed by the estimation of the model parameters easier since the definition of a \emph{common move} gets wider and therefore the identification of these moves by the estimation procedure is easier (see Section \ref{sec:estimation}). Furthermore, we only have to estimate the corresponding tendencies and not the magnitudes of the common move, which might vary between companies.

Note that the KP model assumes stronger dependencies between companies in the same rating class than model \eqref{eq:model}. To see this, let $n_1$ and $n_2$ be two companies in sectors $s_1$ and $s_2$ respectively. In \eqref{eq:KPModel}, due to the independence properties of $\delta_{n_i}^t$ and $\xi_{n_i}^t$ for $i=1,\; 2$, we have
\begin{eqnarray}
\operatorname{Cov}(X_{n_1}^t, X_{n_2}^t) &=& \operatorname{Cov}\left(\delta_{n_1}^t \xi_{n_1}^t+(1-\delta_{n_1}^t)\eta^t_{X_{n_1}^{t-1}}, \delta_{n_2}^t \xi_{n_2}^t+(1-\delta_{n_2}^t)\eta^t_{X_{n_2}^{t-1}} \right) \\
&=& \operatorname{Cov}\left( (1-\delta_{n_1}^t)\eta^t_{X_{n_1}^{t-1}}, (1-\delta_{n_2}^t)\eta^t_{X_{n_2}^{t-1}} \right) \\
&=& \Ed(1-\delta_{n_1}^t) \Ed(1-\delta_{n_2}^t) \Ed(\eta^t_{X_{n_1}^{t-1}}\eta^t_{X_{n_2}^{t-1}}) \\
&-& \Ed(1-\delta_{n_1}^t) \Ed(1-\delta_{n_2}^t) \Ed(\eta^t_{X_{n_1}^{t-1}}) \Ed(\eta^t_{X_{n_2}^{t-1}}) \\
&=& (1-q_{s_1, X_{n_1}^{t-1}}) (1-q_{s_2, X_{n_2}^{t-1}}) \operatorname{Cov}(\eta_{X_{n_1}^{t-1}}^t, \eta_{X_{n_2}^{t-1}}^t).
\end{eqnarray}
Similarly, we get
\begin{eqnarray}
\operatorname{Cov}(X_{n_1}^t, X_{n_2}^t) = (1-q_{s_1, X_{n_1}^{t-1}}) (1-q_{s_2, X_{n_2}^{t-1}}) \operatorname{Cov}(\eta_{n_1}^t, \eta_{n_2}^t)
\end{eqnarray}
for model \eqref{eq:model}. Note that the two above expressions coincide as long as $x_{n_1}^{t-1} \neq x_{n_2}^{t-1}$, i.e. the companies belong to different rating classes at time $t-1$. However, if $x_{n_1}^{t-1} = x_{n_2}^{t-1}$, we trivially have $\operatorname{Corr}(\eta_{X_{n_1}^{t-1}}^t, \eta_{X_{n_2}^{t-1}}^t) = 1$ for the KP model while for \eqref{eq:model}
\begin{equation}
\operatorname{Corr}(\eta_{n_1}^t, \eta_{n_2}^t) \leq 1
\end{equation}
with the exact value depending on the unconditional transition probabilities $P$.

As mentioned above, the second difference is the estimation of model parameters: In \cite{KP07} pairwise correlations between the tendency variables are assumed to be known and the joint distribution of these variables is found by an optimization approach. The probabilities in $Q$ are manually tuned for small models to demonstrate the models flexibility and expressiveness. However, to use the model in practice, numerically tractable estimation routines are needed to obtain realistic parameter estimates for $Q$ and $P_\chi$. We therefore adopt a maximum likelihood approach to estimate the parameters of the model as discussed in the next section.

%%%%%%%%%%%%%%%%%%%%%%%%%%%%%%%%%%%%%%%
% SECTION 3: Estimation of Parameters %
%%%%%%%%%%%%%%%%%%%%%%%%%%%%%%%%%%%%%%%
\section{Maximum Likelihood Approach to Parameter Estimation}
\label{sec:ml}
In this section, we introduce an approach to estimate the parameters of model \eqref{eq:model}, where $P$ is assumed to be known. For a model instance with $M$ non-default rating classes and $S$ industry sectors, there are $S \times M$ unknown parameters in $Q$ and $2^M - (M+1)$ degrees of freedom for the specification of the joint distribution $\chi^t = (\chi_1^t, \ldots, \chi_M^t)$.

Given a set of realizations of the rating process for $N$ firms $x =(x^1, \ldots, x^T) \in \Rd^{N \times T }$ of the $N$ dimensional process $X^t$ for a period of $T$ time steps, we estimate the parameters $Q = (q_{i,j})_{i,j}$ and the joint probability mass function $P_\chi$ for $\chi = (\chi_1, \ldots, \chi_M)$. Since we model rating transitions as coupled Markov chains, $x$ provides us with $N(T-1)$ realizations of the process \eqref{eq:model}, but only with $(T-1)$ (hidden) joint realizations of $\chi = (\chi_1, \ldots, \chi_M)$. It follows that we can analyze every time step separately, but have to take into consideration the joint behavior of moves within each step. In the following, we derive the likelihood function by conditioning.

Because of the Markov property, the likelihood of $x$ given $Q$ and $P_\chi$ is
\begin{eqnarray} \label{ml}
L(x; Q, P_{\chi}) = \prod_{t=2}^{T} \Pd\left( X^t = x^t \mid X^{t-1} = x^{t-1} \right).
\end{eqnarray}
The fact that $\chi^t$ is independent of $X^{t-1}$ yields $P_\chi(\chi^t = \bar{\chi} | X^{t-1} = x^{t-1}) = P_\chi(\chi^t = \bar{\chi})$ and therefore, by the law of total probability, we have
\begin{eqnarray}
\Pd( X^t = x^t \mid X^{t-1} = x^{t-1} ) = \sum_{\bar{\chi} \in \{ 0, 1 \}^M} P_\chi( \chi^t = \bar{\chi} ) \Pd( X^t= x^t| \chi^t = \bar{\chi}, X^{t-1} = x^{t-1})
\end{eqnarray}
where $X^t = (X^t_1, \ldots, X^t_N)$ and $\chi^t=(\chi_1^t, \ldots, \chi_M^t)$.

To calculate the above sum, we divide the companies into groups. Let $I^t(s, m_1, m_2)$ be the number of companies in sector $s$ which move from class $m_1$ to class $m_2$ in period $t$. We start by analyzing these subgroups and fix $t$, $s$, $m_1$ and $m_2$ with $m_1 < m_2$ and $\bar{\chi}_{m_1}=1$. We calculate the probability that the $I^t(s, m_1, m_2)$ companies move from $m_1$ to $m_2$ as a function of the parameters to be estimated. Since $\bar{\chi}_{m_1}=1$, the only possibility for a deterioration $m_1 \to m_2$ to happen is that $\delta_n^t = 1$ for all the corresponding companies. Therefore, the joint probability for these moves is $\left( p_{m_1, m_2}q_{m_1, s} \right)^{I^t(s, m_1, m_2)}.$
%\begin{equation}
%\left( p_{m_1, m_2}q_{m_1, s} \right)^{I^t(s, m_1, m_2)}.
%\end{equation}

Now consider the case that $m_1 \geq m_2$ and $\bar{\chi}_{m_1}=1$. In this case, the corresponding companies could move from $m_1 \to m_2$ either by a realization of $\xi_n^t$ (if $\delta_n^t = 1$) or by a realization of $\eta_n^t$ (if $\delta_n = 0$). Since all the combinations have to be considered, the probability is
\begin{eqnarray}
\sum_{i=0}^{I^t} \binom{I^t}{i} \left( \frac{(1-q_{m_1,s})p_{m_1, m_2}}{p_{m_1}^+} \right)^i  \left( q_{m_1, s} p_{m_1, m_2}\right) ^{I^t-i} = p_{m_1, m_2}^{I^t} \left( \frac{q_{m_1,s}(p_{m_1}^+-1)  +1}{p_{m_1}^+} \right)^{I^t}
\end{eqnarray}
where we abbreviate $I^t(s, m_1, m_2)$ by $I^t$. A similar logic applies for the case $\bar{\chi}_{m_1}=0$.

Splitting up all the moves in period $t$ according to industry sector and rating class, we get by the above argument
\begin{eqnarray}
\Pd( X^t = x^t| \chi^t = \bar{\chi}, X^{t-1} = x^{t-1}) = \prod_{s = 1}^S \prod_{m_1=1}^M \prod_{m_2 = 1}^{M+1} f^t(x, s, m_1, m_2,; Q, P_\chi)
\end{eqnarray}
with
\begin{equation}
f^t(x, s, m_1, m_2,; Q, P_\chi) = \begin{cases} p_{m_1, m_2}^{I^t} \left( \frac{q_{m_1,s}(p_{m_1}^+-1)  +1}{p_{m_1}^+} \right)^{I^t}, & m_1 \geq m_2, \; \bar{\chi}_{m_1} = 1\\
p_{m_1, m_2}^{I^t} \left( \frac{q_{m_1,s}(p_{m_1}^- -1)  +1}{p_{m_1}^-} \right)^{I^t}, & m_1 < m_2, \; \bar{\chi}_{m_1} = 0 \\
p_{m_1, m_2}^{I^t}q_{m_1, s}^{I^t}, & \mbox{otherwise}
\end{cases}
\end{equation}
where $p_m^- = 1- p_m^+$. The likelihood function \eqref{ml} can consequently be written as
\begin{eqnarray}
L(x; Q, P_{\chi}) &=& \prod_{t=2}^{T} \sum_{\bar{\chi} \in \{ 0, 1 \}^M} P_\chi( \chi^t = \bar{\chi} ) \Pd( X^t= x^t| \chi^t = \bar{\chi}, X^{t-1} = x^{t-1} ) \\
&=&\prod_{t=2}^{T} \sum_{\bar{\chi} \in \{ 0, 1 \}^M} P_\chi( \chi^t = \bar{\chi} ) \prod_{s, m_1, m_2} f^t(x, s, m_1, m_2,; Q, P_\chi).
\end{eqnarray}

The above function is clearly non-convex in $P$ and $Q$, and since it consists of mix of sums and products, this problem can not be resolved by a logarithmic transform. Maximizing the likelihood for given data $x$ in the parameters $P_\chi$ and $Q$ amounts to solving the following constrained optimization problem
\begin{eqnarray} \label{genProb}
\begin{array}{llll}
\max_{Q, P_\chi} & \mathcal{L}(x; Q, P_{\chi}) & \\
s.t. 	& q_{m, s} & \in [0,1], & \forall m: 1 \leq m \leq M, \; \forall s: 1\leq s \leq S \\
		& \sum_{\bar{\chi}: \bar{\chi}_m = 1} P_\chi(\bar{\chi}) &= p_m^+, & \forall m: 1 \leq m \leq M \\
		& \sum_{\bar{\chi} \in \set{0,1}^M} P_\chi(\bar{\chi}) &= 1, \\
\end{array}
\end{eqnarray}
where $\Lc$ is the log likelihood function of the model. Note that the constraints in \eqref{genProb} imply that $\sum_{\bar{\chi}: \bar{\chi}_m = 0} P_\chi(\bar{\chi}) + \sum_{\bar{\chi}: \bar{\chi}_m = 1} P_\chi(\bar{\chi})=1$ for all $m$ with $1\leq m \leq M$ and therefore, $\sum_{\bar{\chi}: \bar{\chi}_m = 0} P_\chi(\bar{\chi}) = 1-p_m^+, \; \forall m: 1 \leq m \leq M.$

It turns out that the evaluation of the likelihood is computationally expensive and numerically unstable because of the terms $p_{m_1,m_2}^{I^t}$, which tend to get very small for high values of $I^t$. To obtain a numerically and computationally more tractable version of the \eqref{genProb}, we define
\begin{equation}
\tilde{f}^t(x, s, m_1, m_2,; Q, P_\chi) = \begin{cases} \left( \frac{q_{m_1,s}(p_{m_1}^+-1)  +1}{p_{m_1}^+} \right)^{I^t}, & m_1 \geq m_2, \; \bar{\chi}_{m_1} = 1\\
\left( \frac{q_{m_1,s}(p_{m_1}^- -1)  +1}{p_{m_1}^-} \right)^{I^t}, & m_1 < m_2, \; \bar{\chi}_{m_1} = 0 \\
q_{m_1, s}^{I^t}, & \mbox{otherwise}. \\
\end{cases}
\end{equation}
and derive a concentrated version $\tilde{\Lc}$ of $\Lc$ by replacing $f^t$ by $\tilde{f}^t$. We now define
\begin{equation}
\tilde{\Lc}(x; Q, P_{\chi}) = \sum_{t=2}^T \log\left( \sum_{\bar{\chi} \in \{ 0, 1 \}^M} P_\chi( \chi^t = \bar{\chi} ) \prod_{s, m_1, m_2} \tilde{f}^t(x, s, m_1, m_2,; Q, P_\chi) \right)
\end{equation}
and replace $\Lc$ by $\tilde{\Lc}$ in \eqref{genProb}. The concentrated likelihood is numerically more tractable and computationally less expensive than the original one, while yielding the same parameter estimates as the original formulation (see \ref{app:modLL} for a proof).

%%%%%%%%%%%%%%%%%%%%%%%%%%%%%%%%%%%%%%%%%%%%%%%%%%%%%%
% SECTION 4: Data Description & Parameter Estimation %
%%%%%%%%%%%%%%%%%%%%%%%%%%%%%%%%%%%%%%%%%%%%%%%%%%%%%%
\section{Data Description \& Parameter Estimation}
\label{sec:estimation}
\subsection{Particle Swarm Algorithm}
Since problem \eqref{genProb} is non-convex and the number of parameters to be estimated is too large to employ standard non-convex solvers, we use heuristic global optimization techniques. In particular, we employ a particle swarm algorithm, as described in \citet{HoWo2009a}, to find a local optimum of \eqref{genProb}. The main idea of the algorithm is that finitely many particles move in the feasible region of the problem. In each iteration of the algorithm, the objective value, corresponding to the position of the particle in the feasible region, is evaluated and the particle moves on. Each particle $k$ \emph{knows} its best position $\hat{x}_k$ till then (in terms of the likelihood function) and every particle \emph{knows} the best position $\hat{g}$ ever seen by any particle. The velocity of a particle changes in such a way that it is drawn to $\hat{x}_k$ and $\hat{g}$ to a random degree. Eventually, all the particles will end up close to one another and near to a local optimum of the problem.

In the following, we give a brief description of the algorithm, which follows the ideas in \cite{KennedyEberhart1995}.
\begin{enumerate}
	\item Choose a convergence threshold $\epsilon > 0$ and $K, i^+ \in  \Nd$.
	
	\item Generate random samples $y_k = (Q^k, P_\chi^k)$ for $k=1, \ldots, K$ from the feasible region of \eqref{genProb}. Each sample is the starting point of a \emph{particle}. Set $\hat{y}_k = y_k$ and $v_k = 0$ for all $k=1, \ldots, K$.

	\item Set $\hat{g} \leftarrow \operatorname{argmax}_k \tilde{\Lc}(y_k)$ and $i\leftarrow 0$.

	\item For all particles $y_k$
		\begin{enumerate}
			\item First compute a velocity $v_k$ for the $k$-th particle
			\begin{equation} \label{psaFly}
			v_k \leftarrow c_0 v_k + c_1 r_1 \circ (\hat{y}_k - y_k) + c_2 r_2 \circ (\hat{g} - y_k)
			\end{equation}
			where $c_0$, $c_1$, $c_2$ are fixed constants, $r_1$ and $r_2$ are component-wise uniformly distributed random matrices of appropriate dimension and $\circ$ is the Hadamard (pointwise) matrix multiplication.  The new position of the particle is
			$$ y_k \leftarrow y_k + v_k.$$

			\item If $\tilde{\Lc}(y_k) > \tilde{\Lc}(\hat{y}_k)$ then $\hat{y}_k \leftarrow y_k$.
		\end{enumerate}
	\item If $\tilde{\Lc}(y_k) > \tilde{\Lc}(\hat{g})$ for some $y_k$, then $\hat{g} \leftarrow y_k$. Set $i \leftarrow i+1$.

	\item If $\operatorname{Var}(\tilde{\Lc}(y_1), \ldots, \tilde{\Lc}(y_K))< \epsilon$ or $i\geq i^+$ terminate the algorithm, otherwise go to 4.
\end{enumerate}

Note that in step 4(a) of the above algorithm, a particle may leave the feasible region by violating the constraints on $P_\chi$ or $Q$. In this case, the particle \emph{bounces off the border} and completes its move in the modified direction. For details on sampling from the feasible region as well as on the bouncing of the particles, we refer to \citet{HoWo2009a}.

In our setting the PSA consistently reaches its \emph{final state} after approximately 50 iterations. After this stage $\hat{g}$ does not improve any further and the variance stays constant. Since it is hard to interpret the absolute values of $\tilde{\Lc}$, a meaningful bound for $\epsilon$ can not be found. We therefore exclusively used the criterion $i>i^+$ to terminate the loop.

The algorithm was implemented in MATLAB 2008a. For our calculations, we set $K=800$ and $i^+=100$, which leads to a runtime of several minutes.

\subsection{Data \& Estimates}
The estimation is based on yearly historical rating data for 10166 companies from all over the world as quoted by Standard \& Poor's over a time horizon of $23$ years ($1985$ - $2007$). The data set comprises of $87296$ observations of rating transitions between the $10$ S\&P rating classes from AAA to D (not all the companies are rated over the whole time horizon). To reduce the number of parameters to be estimate, the $10$ rating classes are clubbed in the following way: $1 \leftarrow \set{AAA, AA}$, $2 \leftarrow \set{A,BBB}$, $3 \leftarrow \set{BB, B}$, $4 \leftarrow \set{CCC, CC, C}$ and $5 \leftarrow \set{D}$, i.e. $M = 4$. Additionally, we incorporate information on industry sectors by distinguishing between $6$ industries according to the SIC classification scheme. Table \ref{tab:data} gives an overview of the composition of the sample used for fitting the model.

\begin{table}
\begin{center}
\begin{tabular}{lllllllllll}
\toprule
                & \multicolumn{9}{c}{Rating Class} \\
Sector (Number)    & 1 & 2 & 3 & 4 & 5 & 6 & 7 & 8 & 9 & 10\\
\midrule
Mining \& Constr. (1)                   &  117  &       344   &   550  &   1078  &   924  &    710  &   84   &  16   &   0   &   40  \\
Manufacturing (2)                            &  194  &       777   &  2613  &   2953  &  2697  &   2749  &  208   &  30   &   0   &  135  \\
Tech. \& Utility (3)    &  258  &      1371   &  3914  &   3778  &  1448  &   1494  &  200   &  27   &   3   &  174  \\
Trade (4)                                    &  22   &      148    &  572   &   871   &  777   &   837   &  45    &  1    &  0    &  38   \\
Finance (5)                                  &  2617 &       9856  &  15317 &   12250 &   5806 &    2617 &   466  &    4  &    3  &   212 \\
Services (6)                                 &  10   &      129    &  437   &   807   &  949   &  1171   &  87    &  3    &  0    &  40\\
\bottomrule
\end{tabular}
\end{center}
\caption{\label{tab:data} Number of companies in the sample. Each company is counted in every period in which it has a non-default rating -- defaults are counted only once.}
\end{table}

We estimate the matrix of transition probabilities $P$ by simple counting as
\begin{equation}
P = \left( \begin{array}{cccccc}
    0.9191   & 0.0798  &  0.0009  &  0.0001  &  0.0001 \\
    0.0212   & 0.9428  &  0.0339  &  0.0008  &  0.0013 \\
    0.0039   & 0.0886  &  0.8678  &  0.0244  &  0.0153 \\
    0.0023   & 0.0079  &  0.1759  &  0.6009  &  0.2131 \\
         0   &      0  &       0  &       0  &  1.0000
\end{array} \right).
\end{equation}
Note that in the data set there are transitions from every non-default rating class to every other rating class, and therefore no problems associated with estimated probabilities being $0$ arise. The aggregation of the $10$ rating classes into a reduced set of $5$ classes was chosen after comparing several possible groupings. The current partition strikes a balance between parsimony and expressive power of the model.

The choices above leave us with $2^4-(4+1) + 4\cdot 6= 37$ parameters to be estimated. The estimated matrix $Q$ can be found in Table \ref{tab:estimatedQ}, while the joint probability function for the $\chi$ can be found in Table \ref{tab:estimatedChi}. Along with the estimates of the elements of $Q$ and $\chi$, we provide standard deviations obtained by running the particle swarm algorithm $50$ times with different randomly sampled starting particles. As can be seen by the generally low values for the standard deviations, the algorithm is stable and converges to more or less the same solution in every run.

However, note that the reported standard deviations, measure the noise associated with the random nature of the particle swarm algorithm. In particular, the estimates are not based on asymptotic theory for maximum likelihood estimation. Hence, the standard deviations can be interpreted as a measure of stability of the stochastic solution algorithm applied to solve problem \eqref{genProb} but cannot be used for hypothesis testing.
\begin{table}
\begin{center}
\begin{tabular}{ccccc}
\toprule
                & \multicolumn{4}{c}{Rating Class} \\
Sector    & 1 & 2 & 3 & 4 \\
\midrule
1&0.1974 (0.0008) &   0.0000 (0.0000) &   0.3745 (0.0023) &   1.0000 (0.0001) \\
2&0.0793 (0.0007) &   0.0000 (0.0000) &   0.3205 (0.0015) &   1.0000 (0.0002) \\
3&0.0168 (0.0005) &   0.0000 (0.0000) &   0.0000 (0.0002) &   1.0000 (0.0007) \\
4&0.0000 (0.0001) &   0.0000 (0.0000) &   0.4943 (0.0035) &   1.0000 (0.0002) \\
5&0.1469 (0.0004) &   0.0428 (0.0001) &   0.5068 (0.0013) &   1.0000 (0.0000) \\
6&0.3127 (0.0032) &   0.0000 (0.0000) &   0.4514 (0.0021) &   1.0000 (0.0001) \\
\bottomrule
\end{tabular}
\end{center}
\caption{\label{tab:estimatedQ} Estimated values for the entries of matrix $Q$. The standard deviations reported in brackets are obtained by solving the fitting problem $50$ times and serve as indicator for the noisiness of the PSA, i.e. can not be used for inference.}
% taken from the file results1stRevision-04-Oct-2010 (mod-date: 04.10.2010, 19:40)
\end{table}

\begin{table}
\begin{center}
\begin{tabular}{cccclccccl}
\toprule
\multicolumn{4}{c}{Rating Class} & & \multicolumn{4}{c}{Rating Class}\\
\cmidrule{1-4} \cmidrule{6-9}
1 & 2 & 3 & 4 & Probability & 1 & 2 & 3 & 4 & Probability \\
\midrule
0 & 0 & 0 & 0 &  0.0000 (0.0000) &   0 & 0 & 0 & 1 &  0.0000 (0.0000) \\
1 & 0 & 0 & 0 &  0.0000 (0.0000) &   1 & 0 & 0 & 1 &  0.0000 (0.0000) \\
0 & 1 & 0 & 0 &  0.0000 (0.0016) &   0 & 1 & 0 & 1 &  0.0000 (0.0007) \\
1 & 1 & 0 & 0 &  0.0397 (0.0035) &   1 & 1 & 0 & 1 &  0.0000 (0.0030) \\
0 & 0 & 1 & 0 &  0.0000 (0.0000) &   0 & 0 & 1 & 1 &  0.0000 (0.0000) \\
1 & 0 & 1 & 0 &  0.0000 (0.0000) &   1 & 0 & 1 & 1 &  0.0360 (0.0000) \\
0 & 1 & 1 & 0 &  0.0000 (0.0043) &   0 & 1 & 1 & 1 &  0.0809 (0.0043) \\
1 & 1 & 1 & 0 &  0.1733 (0.0063) &   1 & 1 & 1 & 1 &  0.6701 (0.0062) \\
\bottomrule
\end{tabular}
\end{center}
\caption{\label{tab:estimatedChi} Probability function of $\chi$, i.e. joint distributions of up and down moves in the different rating classes. The first $4$ columns of the above table describe the event, i.e. the combination of up and down moves in the different classes ($1$ being a non-deteriorating move, while $0$ is a deteriorating move), while the last column gives the estimated probability and the standard deviations obtained by solving the fitting problem $50$ times. As it is the case for the standard deviations reported in Table \ref{tab:estimatedQ}, the standard deviations can not be used for inferential statements about the parameters.}
% taken from the file results1stRevision-04-Oct-2010 (mod-date: 04.10.2010, 19:40)
\end{table}

To test the validity of our modeling, we estimated the parameters of the model for the data set restricted to the transitions of the $6499$ US companies in our sample. The results prove to be quite stable: All but one estimated parameter of the restricted model shows an absolute deviation less than 0.02 from the respective values for the full data set. The average absolute deviation is well under 0.01. This can be seen as evidence that the partitioning of companies according to industry sector is more meaningful than the partitioning into geographical subgroups.

Looking at the estimated values for $Q$, we note that $q_{m,s}=1$ for all the entries corresponding to the lowest rating class. This implies that companies \emph{next to default} are mainly influenced by idiosyncratic factors and not by the economic environment. Another interesting finding is that companies belonging to the industry sector \emph{Transportation, Technology and Utility} show a strong dependency on common economic factors, since $q_{m,3} \approx 0$ for all $m:m \leq 3$. In general, rating classes $1$ and $3$ seem to be more affected by common moves than the classes $2$ and $4$.

The estimate of the joint probability mass function of the tendency variables $\chi$ reveal that most combinations of up and down moves are assigned probability zero and that by far the most probable outcome is a non-deteriorating move for all the rating classes. Another entry with positive probability corresponds to the event that the investment grade assets (i.e. the first two classes) make a non-deteriorating move while companies in the other classes face a downward trend. This seems to be a plausible scenario in economically difficult times. A similar interpretation fits to the entry $(1,1,1,0)$. However, slightly surprising there are two entries with positive probabilities which correspond to a downgrading of the highest and the second highest rating class, while the other classes are not downgraded.

In practice, one might want to restrict the number of parameters of the model by introducing tighter constraints on, for example, the entries of the matrix $Q$. As suggested by a referee, we exemplify this by testing whether the plausible restrictions $q_{m_1, s} = q_{m_2, s}$ for all $m_1$, $m_2$ and $s$ lead to a more compact model without losing too much of the statistical quality. Examining the entries in $Q$ does not suggest that the restriction is justified. Fitting the model with the abovementioned restriction reveals that the Bayesian information criterion (BIC) for the restricted model is 43777 while the BIC for the original model is 43401. Hence, the original model is preferable. This is confirmed by a likelihood ratio test, which rejects the restricted model for every reasonable level of significance.

%Another set of possible restrictions reducing model complexity concerns the probabilities of the joint moves represented by $P_\chi$. It seems plausible that the probability of certain joint moves is $0$ or close to zero: as an example, consider the situation where the joint move is such that the highest rating grade deteriorates while lower grades perform non-deteriorating moves.

%%%%%%%%%%%%%%%%%%%%%%%%%%%%%%%
% SECTION 5: Model Comparison %
%%%%%%%%%%%%%%%%%%%%%%%%%%%%%%%
\section{Model Comparison}
\label{sec:comparison}
\subsection{A Benchmark Model}
To assess the quality of the CMC model, we compare it with a model proposed in \cite{McWe06} adapted to our setting. The aforementioned paper discusses GLMM models with latent factors. Debtors are split into buckets $\mathcal{H} = \set{1, \ldots, H}$, which in our case consist of $\mathcal{H} = \set{(m,s): 1\leq m \leq M, \; 1\leq s \leq S}$, i.e. all the combinations of industry sectors and ratings classes. The distribution of rating migrations for debtor $n$ in rating class $m$ and industry sector $s$ at time $t$ is modeled as
\begin{eqnarray} \label{eq:transProb}
\Pd(X_i^t \leq m' \big| \mu, \xi ) = g( \mu_{m,m'} + \phi_m \xi^t_s )
\end{eqnarray}
where $\mu_1, \ldots, \mu_M$ are vectors in $\Rd^{M+1}$ with increasing components, i.e. $\mu_{i,1} \leq \cdots \leq \mu_{i,M+1}$ ($\mu_{i,j}$ refers to the j-th component of the i-th vector). Furthermore, $\xi^t_s$ are random latent systematic factors used for modeling the dependency in rating transitions between debtors of different industrial sectors, while $(\phi_1, \ldots, \phi_M) \in \Rd^M$ are factor loadings, which make this dependency rating class specific. Finally, the function $g:\Rd \to \Rd$ is an arbitrary strictly increasing function which we choose as the \emph{logit} link function, i.e.  $g(x) = (1+\exp(-x))^{-1}$. It is assumed that conditional on the latent factors the rating transitions of the companies are independent.

Inspecting the model, it can be seen that the joint distribution of credit migrations depends on the current ratings as well as the sectoral information. In this sense, the model is similar to the model proposed in this paper. Except for the lack of an autoregressive component in the latent factors, \eqref{eq:transProb} resembles model $(S3)$ proposed in section 4.5 in \cite{McWe06}. In this paper, we did not include a autoregressive term because we are dealing with yearly data as opposed to \cite{McWe06} who analyze quarterly transitions. As a result, there are too few time periods to estimate the parameters of the autoregressive process in a reliable manner: numerical experiments show that inclusions of an autoregressive term dramatically decreases the quality of the estimates. Furthermore, it turns out that estimated autocorrelation is close to zero -- an observation which is plausible when considering yearly data.

To obtain the joint distribution of the rating transitions, we define the migration counts $N^t_{(m,s)} = (N^t_{(m,s),1}, \ldots,  N^t_{(m,s),M+1})$, where $N^t_{(m,s),j}$ is the number of debtors from industry class $s$ with rating $m$ at time $t$ which are in rating class $j$ at the beginning of period $t+1$. We further define $n^t_{(m,s)}$ as the overall number of debtors in rating class $m$ and industry sector $s$ at the beginning of period $t$.

Based on \eqref{eq:transProb}, the joint distribution of the migrations in period $t$ is given by the following multinomial distribution
\begin{equation} \label{eq:multNom}
\Pd(N^t_{(m,s)} = \mathbf{u} |\mu, \xi) = \frac{n^t_{(m,s)}!}{u_1! u_2! \cdots u_{M+1}!} \prod_{l=1}^{M+1} p_{(m,s), l}(\mu, \xi)
\end{equation}
where $\mathbf{u} \in \Nd^{M+1}$ with $u_1 + \cdots + u_{M+1} = n^t_{(m,s)}$ and
\begin{equation}
p_{(m,s), l}(\mu, \xi) = g( \mu_{m,l} + \phi_m \xi^t_s ) - g( \mu_{m,l-1} + \phi_m \xi^t_s )
\end{equation}
with $\mu_{m, 0} = -\infty$ for all $m \in \set{1, \ldots, M}$.

Following \cite{McWe06}, we estimate the model parameters by an application of the Gibbs sampler. To do this, we assign prior distributions to the parameters and unobserved variables in the model. In specific, we assign the independent normal prior $N(0,\omega)$ to the variables $\xi_s^t$, the vague prior $\operatorname{InverseGamma}(0,0)$ to $\omega^2$, an ordered Gaussian prior with covariance matrix $\tau^2 I$ to the vector $\mu$ (where $\tau=100$ is chosen large so as to ensure an non-informative prior), and finally the normal prior $N(0,\tau I)$ to $\phi$. To ensure identifiability of the model, we fix the sign of $\phi_1$ to be $+1$.

When estimating the parameters of the model \eqref{eq:transProb}, we also implicitly estimate the unconditional transition probabilities, i.e. the matrix $P$. However, the CMC model takes the matrix $P$ as an input. Thus, to ensure a fair comparison of the two models, we restrict the parameters $\mu_{m_1, m_2}$ for $1\leq m_1,m_2 \leq M$ in such a way that the unconditional transition probabilities equal $P$ given the parameters $\omega$ and $\phi$.

To fit the model, we iteratively sample from the full conditional distributions of the parameters. Samples from the full conditional distributions are generated using the ARMS algorithm, see \cite{Gilk92}. For the numerical studies, we simulated $5000$ iterations of the Gibbs sampler after an initial burn-in phase of $45000$ iterations.\footnote{The estimation is implemented in R using the \emph{arms()} method in the HI package for sampling from the full conditionals. One run of the Gibbs sampler takes several hours to complete. For a more detailed introduction into the topic of Gibbs sampling see \cite{RoCa04} or in the context of the discussed models \cite{McWe06, HuYu10}.} The results of the estimation along with standard deviations are reported in Table \ref{tab:muEst} and Table \ref{tab:omegaPhi}.\footnote{Note that, as opposed to the Markov Chain model, the standard deviations can be used to make inferential statements on the true values of the parameters.}
\begin{table}
\begin{center}
\begin{tabular}{cllllll}
\toprule
 &   \multicolumn{6}{c}{$m_2$} \\
\cmidrule{2-7}
$m_1$ &   \multicolumn{1}{c}{0} & \multicolumn{1}{c}{1} & \multicolumn{1}{c}{2} & \multicolumn{1}{c}{3} & \multicolumn{1}{c}{4} & \multicolumn{1}{c}{5} \\
\midrule
 1 &   $-\infty$  &  2.5460  &  6.9421  &  8.6172  &  9.7158  & $\infty$ \\ \addlinespace[-7pt]
   &    (0)       &  (0.0233) & (0.0289) & (0.0289) & (0.0297) & (0)   \\
 2 &   $-\infty$  & -3.8586  &  3.3129  &  6.1926  &  6.7005  & $\infty$ \\ \addlinespace[-7pt]
   &   (0)        &  (0.0060) & (0.0059) & (0.0059) & (0.0064) & (0)     \\
 3 &   $-\infty$  & -5.9054  & -2.5644  &  3.5128  &  4.5164 & $\infty$ \\ \addlinespace[-7pt]
   &   (0)        & (0.0621) & (0.0453) &  (0.0453) & (0.0593) & (0) \\
 4 &   $-\infty$  & -6.3538  & -4.8364  & -1.6294 &  1.4457  & $\infty$\\ \addlinespace[-7pt]
   &   (0)        & (0.0577) & (0.0562) & (0.0562) & (0.0286) & (0) \\
\bottomrule
\end{tabular}
\end{center}
\caption{\label{tab:muEst} Parameter estimates as well as standard deviations (in brackets) for the parameters $\mu_{m_1,m_2}$, for $1\leq m_1 \leq M$ and $0\leq m_2 \leq M+1$.}
\end{table}

\begin{table}
\begin{center}
\begin{tabular}{ccccc}
\toprule
$\omega$ & $\phi_1$ & $\phi_2$ & $\phi_3$ & $\phi_4$ \\
\midrule
0.5284 (0.0537) & 1 (0) & 0.4389 (0.0497) & 1.6220 (0.1036) & 1.3698 (0.1353) \\
\bottomrule
\end{tabular}
\end{center}
\caption{\label{tab:omegaPhi} Parameter estimates as well as standard deviations (in brackets) for the parameter $\omega$ and $\phi = (\phi_1, \ldots, \phi_M)^\top$.}
\end{table}

\subsection{Comparing Investment Decisions}
To compare the two models, we set up a portfolio optimization framework, which uses the rating transition process as input. More specifically, we are interested in the risk minimal allocation of capital among a pre-specified set of corporate bonds. For this purpose, we represent the uncertainty about rating transitions of the companies as scenarios. The scenarios are sampled either from the GLMM model \eqref{eq:transProb} or the CMC model \eqref{eq:model} with the parameters as fitted in the last sections.

We base our analysis on 10,000 scenarios from each of the two models for a risk free rate of $\rho = 4\%$. To translate the simulated rating transitions into scenarios for losses, we implement a mark-to-market approach. Consider a scenario $x_n = (x_n^0, \ldots, x_n^T)$ of rating transitions for $T$ years of company $n$, i.e. $x_n$ is a realization of the Markov chain model in \eqref{eq:transProb} or \eqref{eq:model}, where $x_n^0$ is the non-stochastic state of the world at the time the investment decision is taken. Assume, without loss of generality, that the price of each bond is $1$ and all bonds are sold at par at the beginning of the planning horizon and mature at time $T'>T$. Setting $T=1$, the discounted loss of a corresponding bond for company $n$ in sector $s$ for scenario $x_n$ is equal to (cf. \cite{GuFiBh97})
\begin{align}
B_n(x_n) = 1 - c(x_n^0, s, T') &- \mathds{1}_{\set{m: m \leq M}}(x_n^1) \Bigg[ c(x_n^0, s, T') \sum_{i=1}^{T'-1} (1+f(x_n^1, s, i))^{-i} \label{eq:bondValue}\\
&+ (1+f(x_n^1, s, T'-1))^{-(T'-1)} \Bigg], \nonumber
\end{align}
where $\mathds{1}_A$ is the indicator function of the set $A$, $c(m,s,T')$ is the yield of a bond in rating class $m$ and industry sector $s$ with maturity $T'$ at time 0, and finally $f(m, s, i)$ are the forward zero rates of a bond for a company in rating class $m$, sector $s$ and time to maturity $i$ years. The forward zero curves are calculated at the risk horizon of 1 year from the yields $c(m,s,i)$. To estimate the $5$ year yields $c(\cdot)$ of bonds in different sectors depending on the rating classes, we average over $9668$ bond spreads as quoted on the 1/4/2008 by Markit Financial Information Services and add the risk free rate. For parsimony, we assumed that coupons are paid yearly and that the first coupon is paid even if the borrower defaults within the first year.

We start our analysis by evaluating the risk return profile of bonds in different sector/rating class combinations. A summary of this analysis can be found in Table \ref{tab:returns}.
\begin{table}
\begin{center}
\begin{tabular}{lcr@{/}lr@{/}lr@{/}lr@{/}l}
\toprule
&                & \multicolumn{8}{c}{Rating} \\
\cmidrule{3-10}
& Sectors			 &  \multicolumn{2}{c}{1}			 & \multicolumn{2}{c}{2}			 & \multicolumn{2}{c}{3}			 & \multicolumn{2}{c}{4}\\
\midrule
\multirow{6}{*}{CMC} & 1			 &-0.0407&-0.0024			 &-0.0399&0.1951			 &-0.0351&1.0000			 &-0.0728&1.0000\\
                        & 2			 &-0.0414&-0.0111			 &-0.0403&0.2175			 &-0.0429&1.0000			 &0.0991&1.0000\\
                        & 3			 &-0.0404&-0.0088			 &-0.0397&0.2254			 &-0.0504&1.0000			 &0.0721&1.0000\\
                        & 4			 &-0.0401&-0.0044			 &-0.0393&0.2318			 &-0.0431&1.0000			 &0.1494&1.0000\\
                        & 5			 &-0.0497&0.0001			 &-0.0437&0.4211			 &-0.1776&1.0000			 &-0.0710&1.0000\\
                        & 6			 &-0.0392&0.0148			 &-0.0386&0.2701			 &-0.0516&1.0000			 &-0.1408&1.0000\\
\midrule
&                & \multicolumn{8}{c}{Rating} \\
\cmidrule{3-10}
& Sectors			 &  \multicolumn{2}{c}{1}			 & \multicolumn{2}{c}{2}			 & \multicolumn{2}{c}{3}			 & \multicolumn{2}{c}{4}\\
\midrule
\multirow{6}{*}{Gibbs}  & 1			 &-0.0407&-0.0093			 &-0.0384&0.2440			 &-0.0491&1.0000			 &-0.0820&1.0000\\
                        & 2			 &-0.0414&-0.0081			 &-0.0394&0.1991			 &-0.0582&1.0000			 &0.0866&1.0000\\
                        & 3			 &-0.0402&-0.0013			 &-0.0385&0.2206			 &-0.0584&1.0000			 &0.0648&1.0000\\
                        & 4			 &-0.0400&0.0014			 &-0.0385&0.2203			 &-0.0609&1.0000			 &0.1391&1.0000\\
                        & 5			 &-0.0497&-0.0047			 &-0.0406&0.3768			 &-0.1954&1.0000			 &-0.0800&1.0000\\
                        & 6			 &-0.0392&0.0076			 &-0.0372&0.2683			 &-0.0676&1.0000			 &-0.1448&1.0000\\
\bottomrule
\end{tabular}
\caption{\label{tab:returns} Estimated expected losses for the sectors as listed in Table \ref{tab:data} and aggregated rating classes as described in Section \ref{sec:estimation} as well as the corresponding values for $\operatorname{CVaR}_{.99}(B_n)$ for the CMC as well as for the GLMM model, in the format expected losses/$\operatorname{CVaR}_{.99}$.}
\end{center}
\end{table}
Since the GLMM model is calibrated to have the same marginal transition probabilities, the results of the model from \cite{McWe06} are very similar to the respective figures for the CMC model. It turns out that the expected discounted losses as well as the risks in different industry sectors are quite different, with the financial sector (sector 5) emerging as the most attractive, except for companies in rating class $4$.

To compare the implications of the two models in the context of risk management, we set up a simple scenario based asset allocation model. An optimal decision consists of a set of non-negative portfolio weights $w=(w_1, \ldots, w_N) \in \Rd^N$, where $w_i$ represents the percentage of the available capital invested into asset $i$. Adopting the scenario based approach, we solve the following problem using the Conditional Value-at-Risk (CVaR) in the objective function
\begin{equation} \label{eq:CVaRProblem}
\begin{array}{llll}
\min_{w,z} & \multicolumn{3}{l}{\operatorname{CVaR_\alpha}(w^\top B)}\\
s.t. 	& w_n & \in [a_n z_n,b_n z_n], & \forall n: 1\leq n \leq N\\
        & w^\top \mathds{1} & = 1\\
        & \Ed(w^\top B) &\leq e \\
        & z_n &\in \set{0,1}, & \forall n: 1\leq n \leq N
\end{array}
\end{equation}
where $x^\top y = \sum_{n=1}^N x_n y_n$ is the inner product and $\mathds{1} \in \Rd^N$ is the vector consisting of all ones. The random variable $B=(B_1, \ldots, B_N): \Omega \to \Rd^N$ describes the discounted losses of the individual bonds, as calculated in \eqref{eq:bondValue} from the rating transition scenarios. We use finitely many equally probable scenarios to represent $B$. Notice that we use the variables $a_n$, $b_n$ and $z_n$ to ensure that we get reasonable portfolio decisions, i.e. weights which are bounded from above and below, and this in turn makes the above problem a mixed integer problem.

The Conditional Value-at-Risk at level $\alpha \in (0,1)$ (CVaR$_\alpha$) of a random variable $X$ is defined as
\begin{equation}
\operatorname{CVaR}_\alpha(X) = \frac{1}{1-\alpha} \int_\alpha^1 F_X^{-1}(t) dt
\end{equation}
where $F_X$ is the cumulative distribution function of $X$ and $F_X^{-1}$ is its left inverse. In our case, the random variable $X$ is the portfolio loss $w^\top B$ and correspondingly $\operatorname{CVaR}_\alpha(w^\top B)$ is the average loss in the $\alpha$\% of the worst scenarios.

Our choice of CVaR as a risk measure is motivated by its favorable theoretical properties. Since CVaR is a convex risk measure it leads to sensible decisions from an economic point of view, favoring diversification over concentration - a property that for example the Value-at-Risk lacks, see \citet{Pflug2000}. Moreover, CVaR being piecewise linear in the finite scenario setting, makes problem \eqref{eq:CVaRProblem} numerically tractable, see \citet{RockafellarUryasev2000}. Lastly, the fact that CVaR is closely related to the Value-at-Risk (VaR), which plays an important role in the Basel accord. In fact, it is easy to see that CVaR is an upper bound for VaR, making portfolios which have favorable CVaR-characteristics also attractive from a VaR perspective.\footnote{It can even be shown that the Conditional Value-at-Risk is the best conservative approximation of the Value-at-Risk from the class of law invariant convex risk measures, which are continuous from above (see \cite{FoSc04}, Theorem 4.61)}

\begin{figure}
\begin{center}
\includegraphics[scale=0.3]{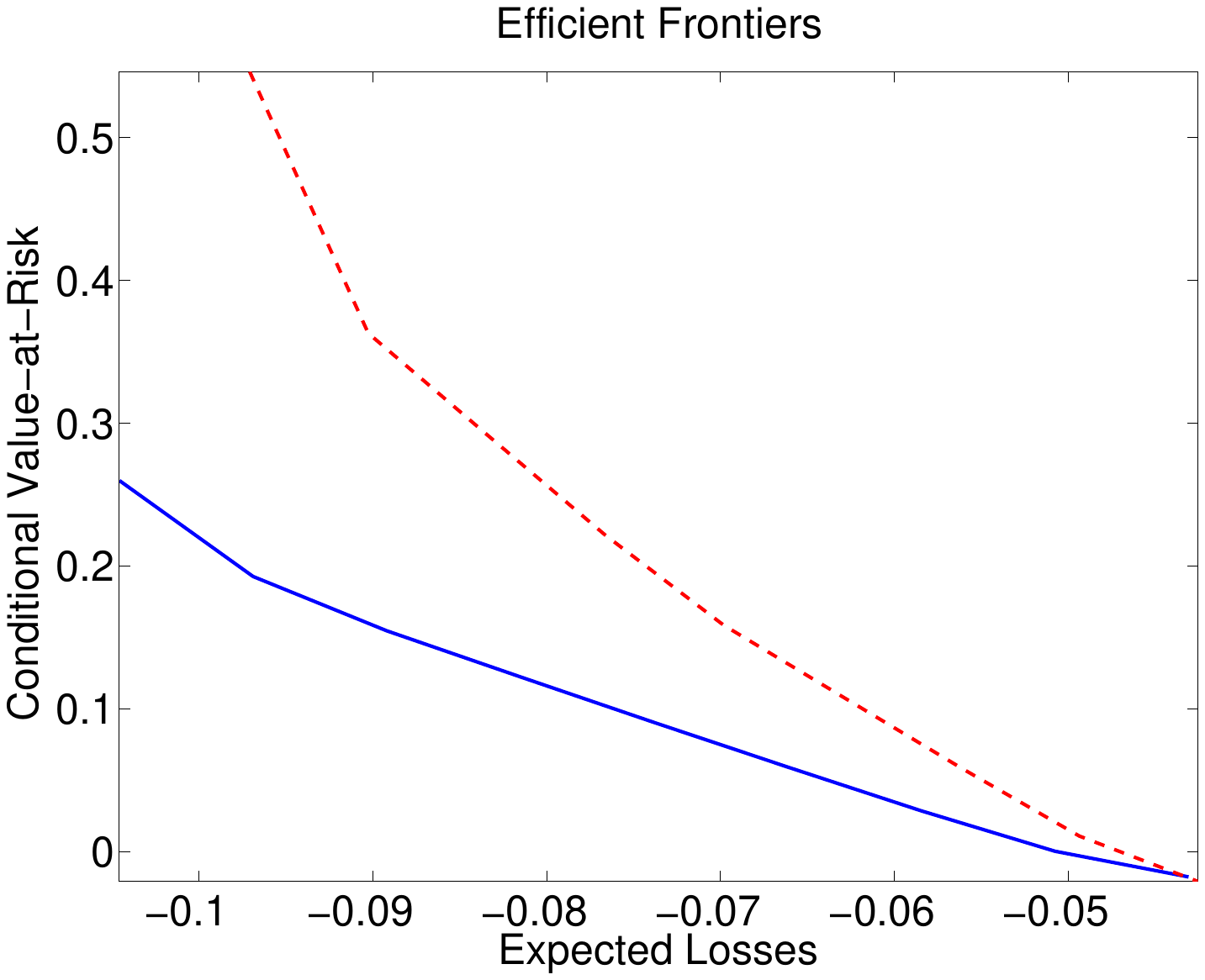}
\end{center}
\caption{\label{fig:effFront} Efficient frontier for problem \eqref{eq:CVaRProblem} calculated with the CMC model (dashed) and the GLMM model (solid).}
\end{figure}

To obtain numerical results, we set up a hypothetical asset universe with one representative bond for each combination of sector and rating. Without loss of generality, we assume that the maturity of all the bonds equals $T'=5$ years and set $T=1$ as discussed above. Furthermore, we set $a_n = 0.01$, $b_n = 0.2$ for all $1 \leq n \leq N$ and $\alpha = 0.99$. To compare how well risks can be hedged for both the models, we solve problem \eqref{eq:CVaRProblem} for varying levels of $e$, spanning the whole range of feasible choices for $e$. The resulting efficient frontier is depicted in Figure \ref{fig:effFront}. Clearly, the scenarios generated by the GLMM model \eqref{eq:transProb} allow for lower risks than the scenarios generated by the CMC models for most of the values of $e$. This is an indication that the scenarios produced by the CMC model exhibit a higher correlation then the results from the GLMM model. Consequently, it is not possible to diversify risks to the same degree as in the GLMM model.

To compare the portfolio decisions produced by the two models, we set $e=-0.04$, i.e. equal to the risk free rate. The results are reported in Table \ref{tab:portfolios}. Interestingly, when using the scenarios from the GLMM model, only assets in the highest rating class are chosen, while the CMC scenarios lead to a more uniform utilization of asset classes with a significant share of the capital in the more risky rating classes. This might be due to the non trivial correlation structure of the variables $\chi_1, \ldots, \chi_m$.
\begin{table}
\begin{center}
\begin{tabular}{cr@{.}lr@{.}lr@{.}lr@{.}lr@{.}lr@{.}l}
\toprule
& \multicolumn{12}{c}{Sector}\\
\cmidrule{2-13}
Rating & \multicolumn{2}{c}{1} & \multicolumn{2}{c}{2} & \multicolumn{2}{c}{3} & \multicolumn{2}{c}{4} & \multicolumn{2}{c}{5} & \multicolumn{2}{c}{6}\\
\midrule
1 &    0&2  &  0&2  &  0&2   & 0&17  &  0&0539   & 0&0393 \\ \addlinespace[-7pt]
    &  0&2  &  0&2  &  0&2   & 0&19  &  0&2      & 0&01 \\
2 &   0&0155 &  0&0165 &  0&0151  & 0&0114 &  0&0145  & 0&0108 \\ \addlinespace[-7pt]
    &  0&0   &  0&0    &  0&0     & 0&0    & 0&0      & 0&0 \\
3 &   0&01 &  0&0108 &  0&0105  & 0&0  &  0&0118   & 0&01 \\ \addlinespace[-7pt]
    & 0&0  &  0&0    &  0&0     & 0&0  &  0&0      & 0&0 \\
4 & 0&0 & 0&0 &     0&0 & 0&0 & 0&0  & 0&0 \\ \addlinespace[-7pt]
    & 0&0 & 0&0 &   0&0 & 0&0 & 0&0  & 0&0 \\
\bottomrule
\end{tabular}
\end{center}
\caption{\label{tab:portfolios} Portfolio composition for the CMC Model on the top and for the GLMM model on the bottom.}
\end{table}

\begin{figure}
\begin{center}
\includegraphics[scale=0.2]{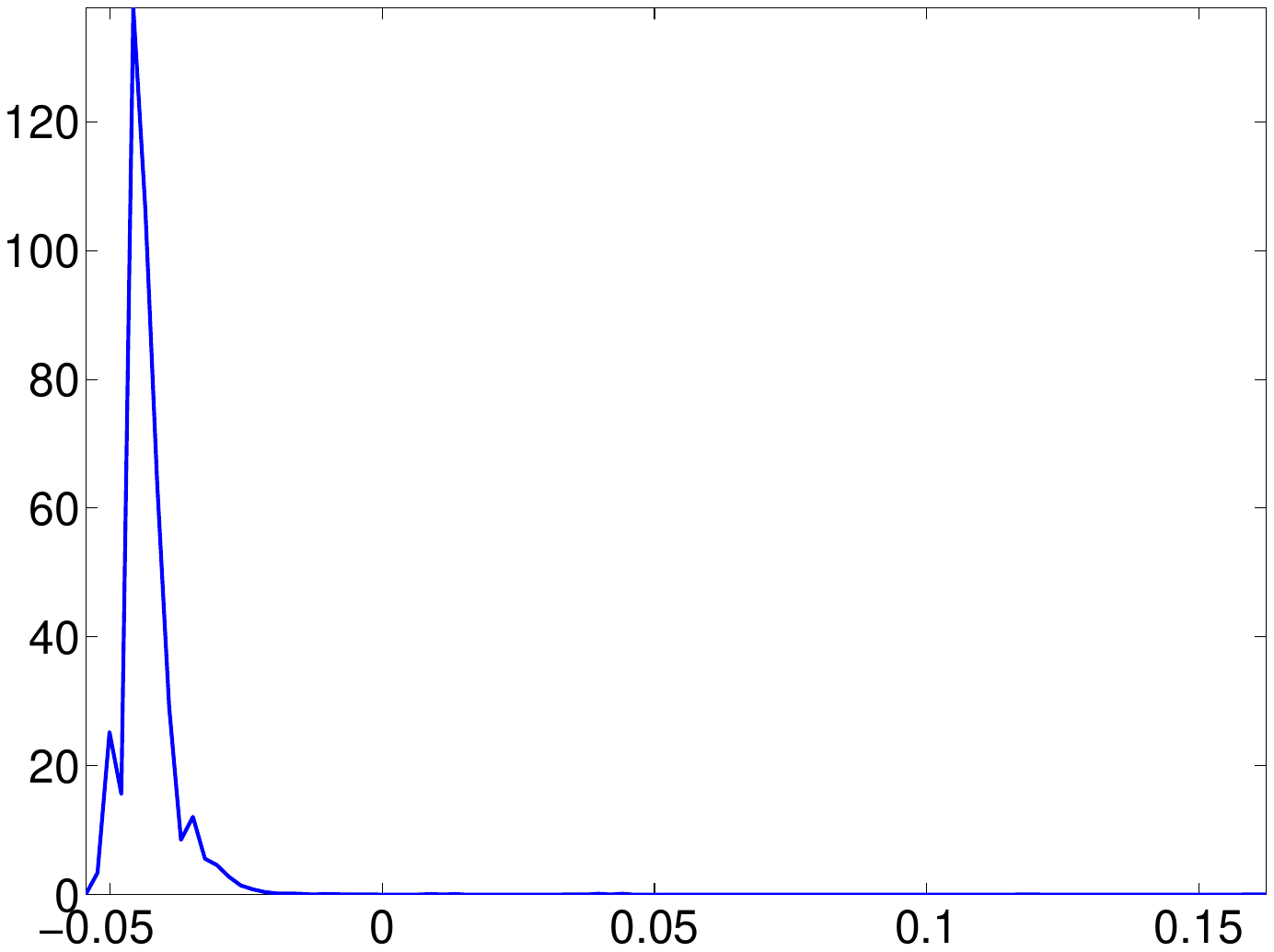} \includegraphics[scale=0.2]{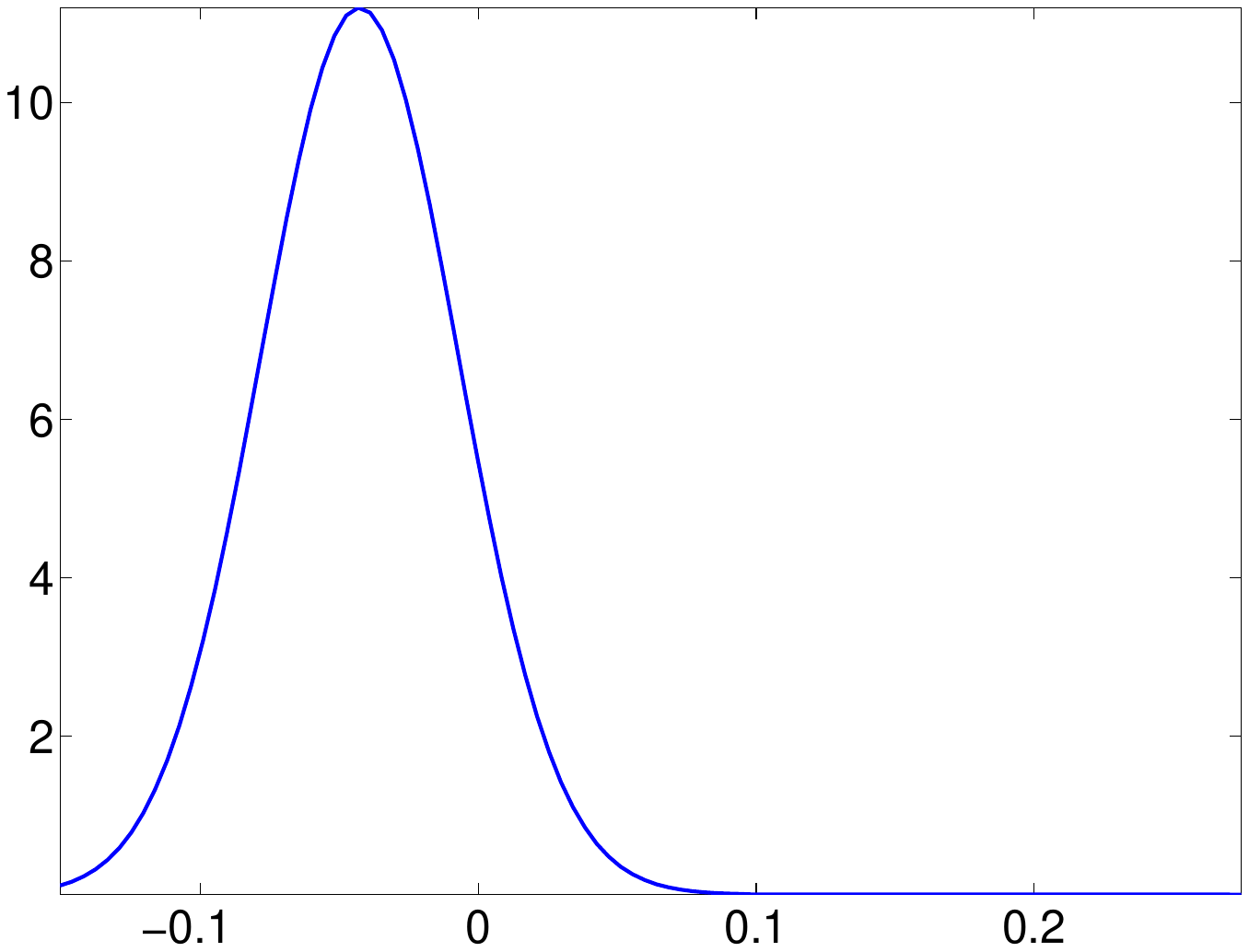}
\end{center}
\caption{\label{fig:returns} Loss distributions of optimal portfolios for $e = -0.04$. On the left losses for the GLMM model are depicted, while the graph on the right show the distribution for the CMC model. The distribution are calculated with the respective loss scenarios and smoothed with a kernel density estimate.}
\end{figure}
The return distributions of the optimal portfolios for $e=-0.04$ are depicted in Figure \ref{fig:returns}. It can be seen that the loss distribution of the optimal portfolio for the GLMM model is much more concentrated and therefore less risky. The return distribution for the CMC model on the other hand has a long right tail and a significant part of the distribution is above 0, i.e. corresponds to losses. This confirms the observation made above that the scenarios representing extreme losses simulated from the CMC model are harder to hedge against. This in turn is another indication to the fact that there are stronger joint migrations in this model.

%%%%%%%%%%%%%%%%%%%%%%%%%
% SECTION 6: Conclusion %
%%%%%%%%%%%%%%%%%%%%%%%%%
\section{Conclusion}
\label{sec:conclusion}
In this paper, we present a coupled Markov chain model for credit rating transitions based on \cite{KP07}. As opposed to the original formulation, our modification lends itself to a maximum likelihood estimation. We derive the likelihood of the model and obtain estimates of the model parameters by solving a simplified but equivalent non-convex optimization problem by heuristic global optimization techniques.

Subsequently, we generate a set of scenarios for joint rating transitions for a set of hypothetical companies and use these to compare the proposed model to a benchmark model from the literature. We find that for the model presented in this paper, there is a stronger dependency between the moves of single debtors. This in turn leads to more conservative portfolio decisions, since extreme risk can not be hedged to the same degree as in the benchmark model.

The flexibility of the approach as well as the computational tractability of large problem instances make the outlined methods interesting for practitioners. %for companies and investors who are faced with the problem of risk management in the context of corporate bonds.

\paragraph{Acknowledgements} The authors want to thank an anonymous referee for careful proofreading and many useful suggestions, which lead to a significant improvement of the paper.

%\newpage
\appendix
\section{Justification of the Modified Likelihood Function}
\label{app:modLL}
Since
\begin{equation}
f^t(x, s, m_1, m_2,; Q, P_\chi) = \tilde{f}^t(x, s, m_1, m_2,; Q, P_\chi) p_{m_1, m_2}^{I^t}
\end{equation}
and $p_{m_1, m_2}$ are fixed parameters not affected by the decision variables $Q$ and $P_\chi$, it is possible to \emph{concentrate out} the terms $p_{m_1, m_2}^{I^t}$ without changing the optimizer of problem \eqref{genProb}. In detail:
\begin{eqnarray}
& &\Lc(x; Q, P_{\chi}) = \sum_{t=2}^T \log\left( \sum_{\bar{\chi} \in \{ 0, 1 \}^M} P_\chi( \chi^t = \bar{\chi} ) \prod_{s, m_1, m_2} f^t(x, s, m_1, m_2; Q, P_\chi) \right) \\
&=& \sum_{t=2}^T \log\left( \sum_{\bar{\chi} \in \{ 0, 1 \}^M} P_\chi( \chi^t = \bar{\chi} ) \prod_{s, m_1, m_2} p_{m_1, m_2}^{I^t} \prod_{s, m_1, m_2} \tilde{f}^t(x, s, m_1, m_2; Q, P_\chi) \right) \\
&=& \sum_{t=2}^T \log\left( \prod_{s, m_1, m_2} p_{m_1, m_2}^{I^t} \sum_{\bar{\chi} \in \{ 0, 1 \}^M} P_\chi( \chi^t = \bar{\chi} ) \prod_{s, m_1, m_2} \tilde{f}^t(x, s, m_1, m_2; Q, P_\chi) \right) \\
&=& \sum_{t=2}^T \log\left( \sum_{\bar{\chi} \in \{ 0, 1 \}^M} P_\chi( \chi^t = \bar{\chi} ) \prod_{s, m_1, m_2} \tilde{f}^t(x, s, m_1, m_2; Q, P_\chi) \right) \\
&+& \sum_{t=2}^T \log\left( \prod_{s, m_1, m_2} p_{m_1, m_2}^{I^t} \right).
\end{eqnarray}
However, the last term does not depend on the decision variables but only on the data, i.e. is a constant in the optimization problem which can be omitted without changing the optimal solution.

\bibliographystyle{plainnat}
\bibliography{cmc}

\end{document}